\documentclass[twocolumn,prd,amsmath,amssymb,showpacs]{revtex4}
\usepackage{graphicx}

\begin{document}

\title{Testing Modified Gravity (MOG) with gas-dominated galaxies}

\author{J. W. Moffat$^{\star,\dagger}$ and
Viktor T. Toth$^\star$\\~\\
{\rm
\footnotesize
$^\star$Perimeter Institute for Theoretical Physics, Waterloo, Ontario N2L 2Y5, Canada\\
$^\dagger$Department of Physics and Astronomy, University of Waterloo, Waterloo, Ontario N2L 3G1, Canada}}

\begin{abstract}
We compare the MOG and MOND predictions of the Tully-Fisher relationship to observations using gas-rich galaxies for which the baryonic mass can be estimated accurately. We find that both theories are in good agreement with the data. Future observations of large, gas-rich galaxies may offer a means to distinguish between these two theories.
\end{abstract}


\maketitle

MOG (MOdified Gravity \cite{Moffat2006a,Moffat2007e}), is a covariant scalar-tensor-vector theory of gravity (STVG) in which the standard gravitational attraction due to the metric tensor is supplemented by a {\em repulsive} Yukawa-type force of finite range and a variable gravitation constant. Phenomenologically, MOG predicts Newtonian gravity at the short range; Newtonian-like gravity with an enhanced value of the gravitational constant at infinity; and an intermediate transitional region. The magnitude of the enhanced gravitational constant at infinity and the range where the transitional region is located are determined by scalar fields. In the case of the spherically symmetric, static vacuum solution, these scalar fields are sourced by the gravitating mass.

For galaxies modeled using MOG, even in absence of exotic dark matter, the rotation velocities in the transitional region follow an approximately flat rotation curve. The relationship between the source mass $M$ and the flat rotational velocity $v_f$ obeys the Tully-Fisher law. It is notable that the Tully-Fisher law emerges from the theory and thus amounts to a prediction; it is not incorporated into MOG by design \cite{Moffat2007e}. In the Tully-Fisher formula, $v_f\propto M^\alpha$, the value of the exponent is $\alpha=4$ for very large and very small masses (albeit for very small masses, the transitional region is non-existent, and flat rotation curves, e.g., for the solar system, are not predicted) and $2\le\alpha\le 4$ for masses in the intermediate range. Notably, many galaxies fall into this intermediate range, with $v_f\sim 100$~km/s.

In past studies \cite{Brownstein2006a,Brownstein:2009gz} the MOG acceleration law was used to match galaxy photometric data against rotational velocity profiles. It was found that the rotational velocities often closely followed the photometric curve (see, in particular, section 4.2 of \cite{Brownstein:2009gz}), and this behavior, which is difficult to mimic using a cold dark matter halo, was easily fitted using MOG. In these fits, the mass-to-light ration $\Upsilon$ was used as a fitted parameter; reasonable values of $\Upsilon$ were obtained, consistent with the known baryonic content of the galaxies studied. However, large uncertainties remain as the amount of baryonic mass present in most galaxies is not well known.

As an alternative, the use of gas-rich galaxies was recently proposed \cite{McGaugh2009,McGaugh2010} as a novel test of the baryonic Tully-Fisher relationship and of alternate gravity theories, notably Modified Newtonian Dynamics (MOND, \cite{Milgrom1984}). Unlike the stellar mass $M_\star$, which is difficult to ascertain and the results are heavily model-dependent, the mass of gas $M_g$ in a galaxy can be directly measured. Therefore, the uncertainty of the total baryonic mass, $M_b=M_\star+M_g$, will be relatively small for galaxies in which $M_g>M_\star$.

A sample of gas-dominated galaxies indeed confirms the baryonic Tully-Fisher relationship $M_b\propto v_f^\alpha$ ($\alpha\simeq 4$) between $M_b$ and the flat rotational velocity $v_f$. Remarkably, the result is also well predicted by MOND \cite{McGaugh2011}, leading to suggestions to seek a more complete gravitational theory that contains MOND in the weak field limit.

MOG presents another alternative. Like MOND, MOG also makes {\em a priori} predictions of galaxy rotation curves that are in good agreement with observation. On the other hand, MOG is a fully relativistic, covariant theory of gravity from the outset, and it can fit observations ranging from the laboratory to cosmological scales without the use of exotic dark matter. In particular, in addition to galaxy rotations, MOG fits galaxy cluster velocity data \cite{Brownstein2006b}, the merging clusters 1E0657-558 (the ``Bullet'' Cluster) \cite{Brownstein2007}, and cosmological observations \cite{Moffat2007e}.

As neither MOND nor MOG have additional adjustable parameters other than the global constants of the theory and mass estimates, these galaxies provide strong constraints on both theories. The relationship between the mass $M_\mathrm{MOG}$ and the rotational velocity $v_f$ in the flat part of the rotation curve under MOG is described by the the equation
\begin{equation}
v_f^2=\left[G_N+(G_\infty-G_N)\frac{M_\mathrm{MOG}}{(\sqrt{M_\mathrm{MOG}}+E)^2}\right]D\sqrt{M_\mathrm{MOG}},
\end{equation}
where $G_N$ is Newton's constant of gravitation, $G_\infty=20G_N$, $D=6250 M_\odot^{1/2}\mathrm{kpc}^{-1}$, and $E=25000 M_\odot^{1/2}$ \cite{Moffat2007e}. The equation can be solved for $M_\mathrm{MOG}$ as a function of $v_f$.

For comparison, the corresponding relationship between the MOND mass $M_\mathrm{MOND}$ and $v_f$ is
\begin{equation}
M_\mathrm{MOND}=\frac{v_f^4}{G_Na_0},
\end{equation}
with $a_0=1.2\times 10^{-10}$~m/s$^2$.

\begin{figure}[ht]
\begin{center}
\includegraphics[width=\linewidth]{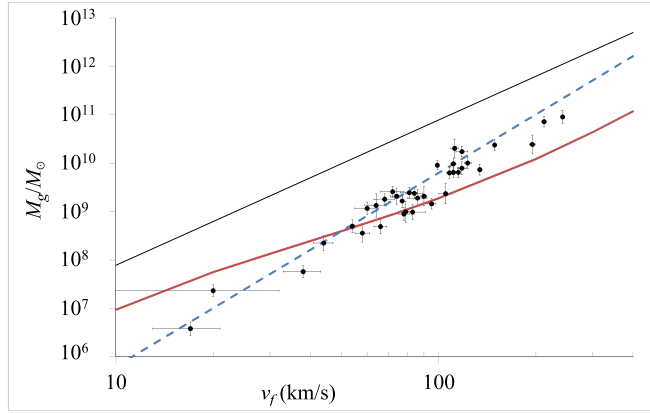}
\caption{\label{fig:MvsVf}Comparison of the MOG (red solid line) and MOND (blue dashed line) predictions to a sample of gas-rich galaxies for which the baryonic mass and flat rotational velocities are known with good accuracy. Note that neither curves are fits: they are predictions obtained from theories with global constants and no per-galaxy fitted parameters. For reference, the ``nominal $\Lambda$CDM'' estimate from \cite{McGaugh2011} is also shown (topmost black solid line).}
\end{center}
\end{figure}

We compared these two predictions against a set of 36 gas-dominated galaxies that was used in \cite{McGaugh2009}. The results are shown in Figure~\ref{fig:MvsVf}.

Our calculations confirm the result reported in \cite{McGaugh2011}: MOND fits the data well. As anticipated, MOG also provides a good fit. Remarkably, qualitative differences exist between the MOG and MOND fits, which may one day be used to distinguish between these theories.

For MOND, the exponent in the Tully-Fisher relationship is fixed at $\alpha=4$. In the case of MOG, the situation is more nuanced. In this theory, the value of the gravitational constant at infinity (i.e., far from the source) increases for larger masses, and may be as high as 20 times the Newtonian value. In the intermediate mass range, where this increase occurs, the value of the Tully-Fisher exponent is reduced, as a proportionately smaller increase in mass is required to yield a given rotational velocity. This explains the elongated ``S''-shape of the MOG curve shown in Figure~\ref{fig:MvsVf}. That the Tully-Fisher exponent may not be exactly 4, and that its average value for a given galaxy sample may be less has already been established (see, e.g., \cite{Brownstein2006a,Brownstein:2009gz}).

Visual examination of Figure~\ref{fig:MvsVf} suggests that both theories are consistent with the data, but the difference between the straight line of MOND and the elongated ``S'' shape of MOG is evident. Visual inspection of a log-log curve and data points with variable error bars can be misleading, however. We obtained a quantitative measure of the fit by calculating a weighted $\chi^2$ statistic for both fits, comparing estimated masses (not logs!) to observation. We constructed weights by treating the uncertainties of the observed mass estimate and the theoretical mass estimate derived using both MOG and MOND from the flat rotational velocity as independent errors, and obtained
\begin{equation}
\chi^2_\mathrm{MOG}/\mathrm{d.f.}=7.98,\quad\chi^2_\mathrm{MOND}/\mathrm{d.f.}=6.69.
\end{equation}
\vskip -3pt

Although the result favors MOND slightly \footnote{Curiously, this is due entirely to a single galaxy in the survey, DDO210, the Aquarius Dwarf. If we remove this galaxy from the data set, MOG is favored. This galaxy is somewhat peculiar as it is one of the few galaxies known to display a blueshift.}, it demonstrates that MOG is a viable modified theory of gravity, offering a possible explanation for galaxy dynamics without having to invoke an exotic dark matter component of unknown constitution. On the other hand, the MOG and MOND predictions differ from one another markedly for very large and very small galaxies. While very small galaxies may be somewhat controversial (e.g., it is difficult to find small galaxies that are unambiguously free of perturbations and tidal disruptions \cite{KroupaDSph1997,KroupaDSph2011}), rotation curves for large, gas-rich galaxies may offer a means to distinguish these two theories in the future.

\section*{Acknowledgments}

The research was partially supported by National Research Council of Canada. Research at the Perimeter Institute for Theoretical Physics is supported by the Government of Canada through NSERC and by the Province of Ontario through the Ministry of Research and Innovation (MRI).

\bibliography{refs}
\bibliographystyle{apsrev}

\end{document}